\documentclass[a4paper,aps,prd,10pt,preprintnumbers,showpacs,twocolumn,superscriptaddress,nofootinbib,amsmath,amssymb]{revtex4-1}
\usepackage[dvips]{graphics}
\usepackage[utf8]{inputenc}
\usepackage[T1]{fontenc}
\usepackage{cmap}

\def\m{\overline{m}}
\def\a{\widetilde{\alpha}}

\def\M{{\cal M}}
\def\Q{Q}
\def\LE{\overline{\Lambda}}
\def\QE{\overline{Q}}

\begin{document}

\title{Black holes in the four-dimensional Einstein-Lovelock gravity}

\author{R. A. Konoplya}\email{roman.konoplya@gmail.com}
\affiliation{Research Centre for Theoretical Physics and Astrophysics, Institute of Physics, Silesian University in Opava, Bezručovo nám. 13, CZ-74601 Opava, Czech Republic}
\affiliation{Peoples Friendship University of Russia (RUDN University), 6 Miklukho-Maklaya Street, Moscow 117198, Russian Federation}

\author{A. Zhidenko}\email{olexandr.zhydenko@ufabc.edu.br}
\affiliation{Research Centre for Theoretical Physics and Astrophysics, Institute of Physics, Silesian University in Opava, Bezručovo nám. 13, CZ-74601 Opava, Czech Republic}
\affiliation{Centro de Matemática, Computação e Cognição (CMCC), Universidade Federal do ABC (UFABC),\\ Rua Abolição, CEP: 09210-180, Santo André, SP, Brazil}

\begin{abstract}
A $(3+1)$-dimensional Einstein-Gauss-Bonnet theory of gravity has been recently formulated in [D.~Glavan and C.~Lin, Phys.\ Rev.\ Lett.\ {\bf 124}, 081301 (2020)] which is different from the pure Einstein theory, i.e., bypasses the Lovelock's theorem and avoids Ostrogradsky instability. The theory was formulated in $D > 4$ dimensions and its action consists of the Einstein-Hilbert term with a cosmological constant, while the Gauss-Bonnet term multiplied by a factor $1/(D-4)$. Then, the four-dimensional theory is defined as the limit $D \to 4$. Here we generalize this approach to the four-dimensional Einstein-Lovelock theory and formulate the most general static $4D$ black-hole solution allowing for a $\Lambda$-term (either positive or negative) and the electric charge $Q$. As metric functions cannot be found in a closed form in the general case, we develop and share publicly the code which constructs the metric functions for every given set of parameters.
\end{abstract}

\maketitle

\section{Introduction}
The Lovelock theorem states that there is no other tensor, which is divergence free, symmetric, and concomitant of the metric tensor and its derivatives in four dimensions, except for the metric tensor and the Einstein tensor \cite{Lovelock}. Therefore, it was concluded that the appropriate vacuum equations in four dimensions are the Einstein equations with the cosmological term. In higher than four-dimensional spacetimes, the most general theory of gravity is generalized by adding higher curvature corrections \cite{Lovelock} to the Einstein action. In five- and six-dimensional cases, the addition term is the (quadratic in curvature) Gauss-Bonnet term, while for higher number of spacetime dimensions $D$, higher curvature terms are required. These are described by the Einstein-Lovelock gravity \cite{Lovelock} for arbitrary $D$. At the same time, higher curvature corrections appear in the low-energy limit of string theory. This induced considerable interest to Einstein-Gauss-Bonnet and Einstein-Lovelock theories, and, especially to black-hole solutions in these theories \cite{Boulware:1985wk,Wheeler,Wiltshire:1985us,Cai:2001dz}.

Higher-dimensional black hole in Einstein-Gauss-Bonnet gravity and its Lovelock generalization were studied in a great number of papers. Such black holes show a number of peculiar properties which are not appropriate for usual Tangherlini black holes \cite{Tangherlini:1963bw}. Thus, when the coupling constants are not small enough, the black holes are unstable and instability develops at high multipoles numbers   \cite{DottiGleiser,Konoplya:2017lhs,Konoplya:2017zwo,Yoshida:2015vua,Takahashi:2011qda,Konoplya:2008ix,Cuyubamba:2016cug,Takahashi:2012np,Takahashi:2010ye}, the lifetime of even slightly Gauss-Bonnet corrected black hole is much longer and the evaporation rate is much smaller \cite{Konoplya:2010vz}, the eikonal gravitational quasinormal modes break down the correspondence between the eikonal quasinormal modes and the null geodesics \cite{Cardoso:2008bp,Konoplya:2017wot}.

Recently, in \cite{Glavan:2019inb}, it has been proposed the way to bypass the conclusions of Lovelock's theorem by performing a dimensional regularization of Gauss-Bonnet equations and obtain a four-dimensional metric theory of gravity with diffeomorphism invariance and second-order equations of motion. The approach has been formulated in $D > 4$ dimensions and its action consists of the Einstein-Hilbert term with a cosmological constant, while the Gauss-Bonnet term multiplied by a factor $1/(D-4)$. Then, the four-dimensional theory is defined as the limit $D \to 4$ of the higher-dimensional theory after the appropriate rescaling of the coupling constant $\alpha$. The properties of black holes in this novel theory, such as (in)stability, quasinormal modes, and shadows, were considered in \cite{Konoplya:2020bxa,Konoplya:2020juj}, while the innermost circular orbits were analyzed in \cite{Guo:2020zmf}. The generalization to the charged black holes and an asymptotically anti-de Sitter (AdS) and de Sitter (dS) cases  in the $4D$  Einstein-Gauss-Bonnet theory was considered in \cite{Fernandes:2020rpa}. Various aspects of the new approach to regularization in the Einstein-Gauss-Bonnet theory have been recently considered in \cite{Fernandes:2020rpa,Singh:2020xju,Zhang:2020qam,Ghosh:2020vpc,HosseiniMansoori:2020yfj,Lu:2020iav,Konoplya:2020ibi}.

Here we generalize this idea for the arbitrary Lovelock theory, that is, we formulate the four-dimensional regularized Einstein-Lovelock theory in a similar way as it was done in \cite{Glavan:2019inb} for the Einstein-Gauss-Bonnet case. Further we discuss static charged black objects in this Einstein-Lovelock theory, allowing for a (either positive or negative) cosmological constant curvature. Thus, among other solutions, we find the metrics describing four-dimensional, charged asymptotically flat, dS and AdS black holes in the theory with higher curvature corrections, and a number of coupling constants. We consider the position of the event horizon for all the branches and the Hawking temperature of the solutions. As in the general case, the metric function cannot be represented in a unique closed form, but is represented by some solutions of the algebraic equations; we also develop and share the Mathematica\textregistered{} code which allows one to construct the metric function for every fixed set of physical parameters, such as values of the coupling constants, charge, mass, $\Lambda$-term, and the number of spacetime dimensions.

Our paper is organized as follows. In Sec.~\ref{sec:Lovelock} we propose the regularized four-dimensional Einstein-Lovelock theory. Sec.~\ref{sec:regularized} discusses static black objects in this theory. In Sec.~\ref{sec:horizons} we discuss the structure of the event horizon and the Hawking temperature of these objects. Sec.~\ref{sec:metric} is devoted to the subclass of compact objects (black holes). Finally, in the Conclusion, we summarize the obtained results and mention an open problem of stability.

\section{Lovelock theory}\label{sec:Lovelock}
The Lagrangian density of the Einstein-Lovelock theory has the form
\cite{Lovelock}
\begin{eqnarray}\label{Lagrangian}
  \mathcal{L} &=& -2\Lambda+\sum_{m=1}^{\m}\frac{1}{2^m}\frac{\alpha_m}{m}
  \delta^{\mu_1\nu_1\mu_2\nu_2 \ldots\mu_m\nu_m}_{\lambda_1\sigma_1\lambda_2\sigma_2\ldots\lambda_m\sigma_m}\,\\\nonumber
  &&\times R_{\mu_1\nu_1}^{\phantom{\mu_1\nu_1}\lambda_1\sigma_1} R_{\mu_2\nu_2}^{\phantom{\mu_2\nu_2}\lambda_2\sigma_2} \ldots R_{\mu_m\nu_m}^{\phantom{\mu_m\nu_m}\lambda_m\sigma_m},
\end{eqnarray}
where
$$\delta^{\mu_1\mu_2\ldots\mu_p}_{\nu_1\nu_2\ldots\nu_p}=\det\left(
\begin{array}{cccc}
\delta^{\mu_1}_{\nu_1} & \delta^{\mu_1}_{\nu_2} & \cdots & \delta^{\mu_1}_{\nu_p} \\
\delta^{\mu_2}_{\nu_1} & \delta^{\mu_2}_{\nu_2} & \cdots & \delta^{\mu_2}_{\nu_p} \\
\vdots & \vdots & \ddots & \vdots \\
\delta^{\mu_p}_{\nu_1} & \delta^{\mu_p}_{\nu_2} & \cdots & \delta^{\mu_p}_{\nu_p}
\end{array}
\right)$$
is the generalized totally antisymmetric Kronecker delta, $R_{\mu\nu}^{\phantom{{\mu\nu}}\lambda\sigma}$ is the Riemann tensor, $\alpha_1=1/8\pi G=1$, and $\alpha_2,\alpha_3,\alpha_4,\ldots$ are arbitrary constants of the theory.

The Euler-Lagrange expression, corresponding to the Lagrangian density (\ref{Lagrangian}) is defined as follows:
$$E^{\mu\nu}\equiv\frac{\partial\sqrt{-g}\mathcal{L}}{\partial g_{\mu\nu}}-\frac{\partial}{\partial x^{\lambda}}\frac{\partial\sqrt{-g}\mathcal{L}}{\partial g_{\mu\nu,\lambda}}+\frac{\partial^2}{\partial x^{\lambda}\partial x^{\sigma}}\frac{\partial\sqrt{-g}\mathcal{L}}{\partial g_{\mu\nu,\lambda\sigma}}.$$
In \cite{Kofinas:2007ns} it was shown that this expression has the following form:
\begin{eqnarray}\label{Lovelock}
  \frac{E^{\mu}_{\nu}}{\sqrt{-g}} &=& \Lambda\delta^{\mu}_{\nu}-\sum_{m=1}^{\m}\frac{1}{2^{m+1}}\frac{\alpha_m}{m}
  \delta^{\mu\mu_1\nu_1\mu_2\nu_2 \ldots\mu_m\nu_m}_{\nu\lambda_1\sigma_1\lambda_2\sigma_2\ldots\lambda_m\sigma_m} \\\nonumber
&& \times R_{\mu_1\nu_1}^{\phantom{\mu_1\nu_1}\lambda_1\sigma_1} R_{\mu_2\nu_2}^{\phantom{\mu_2\nu_2}\lambda_2\sigma_2} \ldots R_{\mu_m\nu_m}^{\phantom{\mu_m\nu_m}\lambda_m\sigma_m}\,.
\end{eqnarray}

The antisymmetric tensor is nonzero only when the indices $\mu,\mu_1,\nu_1,\mu_2,\nu_2,\ldots\mu_m,\nu_m$ are all distinct. Thus, the general Lovelock theory is such that $2\m$ is smaller than the number of spacetime dimensions $D$. In particular, in $D=4$ spacetime the Lovelock theorem implies that $\m=1$, which is equivalent to the Einstein theory \cite{Lovelock}. When $D=5$ or $6$, $\m=2$, implying that the correction to the Einstein action is quadratic in curvature (in the Gauss-Bonnet form) and a new appropriate nonvanishing coupling constant $\alpha_2$ appears.

Assuming a maximally symmetric solution of the $D$-dimensional theory, we consider
\begin{equation}\label{MSR}
  R_{\mu\nu}^{\phantom{\mu\nu}\lambda\sigma}=\frac{2\Lambda_e}{(D-1)(D-2)}\delta^{\lambda\sigma}_{\mu\nu},
\end{equation}
where $\Lambda_e$ is the effective cosmological constant.

Substituting (\ref{MSR}) into (\ref{Lovelock}) we find that $E^{\mu}_{\nu}=0$ yields
\begin{equation}
\Lambda=\sum_{m=1}^{\m}\frac{\alpha_m}{2m}\frac{(D-1)!}{(D-2m-1)!}\left(\frac{2\Lambda_e}{(D-1)(D-2)}\right)^m\,.
\end{equation}

It follows that the effective cosmological constant gets corrections due to the Lovelock terms,
\begin{equation}\label{Leff}
\frac{2(\Lambda-\Lambda_e)}{(D-1)(D-2)}=\sum_{m=2}^{\m}\a_m\left(\frac{2\Lambda_e}{(D-1)(D-2)}\right)^m,
\end{equation}
where, following \cite{Konoplya:2017lhs}, we have introduced
\begin{equation}
\a_m=\frac{\alpha_m}{m}\frac{(D-3)!}{(D-2m-1)!}=\frac{\alpha_m}{m}\prod_{p=1}^{2m-2}(D-2-p).
\end{equation}

We can see that $\a_2=0$ for $D=4$, $\a_3=0$ for $D=4,5,6$, $\a_4=0$ for $D=4,5,6,7,8$ etc. Therefore, taking the limit of $\alpha_m$ such that $\a_m$ remain constant, we obtain the regularized theory, which generalizes the approach of \cite{Glavan:2019inb}.
Indeed, in the case of Einstein-Gauss-Bonnet theory ($\m=2$), the above equation reads
\begin{equation}
\alpha_2 = \frac{2 \a_2}{(D-3)(D-4)}.
\end{equation}
Then, taking the limit $D\to4$ leads to
\begin{equation}
\alpha_2 \to \frac{2 \a_2}{D-4}.
\end{equation}
Thus, our units differ by a factor $2$ from those used in \cite{Glavan:2019inb} and coincides with the units of \cite{Fernandes:2020rpa}. It is worthwhile noticing that prior to \cite{Glavan:2019inb} the dimensional regularization of this kind was suggested in an unpublished work by Y.~Tomozawa~\cite{Tomozawa:2011gp}, where it was shown that the dimensional regularization of the first-order quantum corrections to gravity indeed corresponds to appearing of the coupling $\a_2$ in the four-dimensional theory. It is natural to expect that higher Lovelock orders should describe the corresponding higher-order quantum corrections.

Although the Lagrangian (\ref{Lagrangian}) diverges in this limit, we conclude that no singular terms appear in the Einstein-Lovelock equations for any $D\geq3$.

In particular, for $D=4$, Eq.~(\ref{Leff}) reads
\begin{equation}\label{Lseries}
\frac{\Lambda-\Lambda_e}{3}=\sum_{m=2}^{\m}\a_m\left(\frac{\Lambda_e}{3}\right)^m=\a_2\frac{\Lambda_e^2}{9}+\a_3\frac{\Lambda_e^3}{27}\ldots.
\end{equation}

The Einstein-Gauss-Bonnet theory (corresponding to $\m=2$) implies the existence of two branches \cite{Glavan:2019inb},
$$\Lambda_e=\frac{3}{2\a_2}\left(-1\pm\sqrt{1-\frac{4\Lambda\a_2}{3}}\right).$$
Only the ``+'' branch is perturbative in $\a_2$, while for the ``-'' branch $\Lambda_e$ diverges in the limit $\a_2 \to 0$.

Similarly, higher-order Lovelock terms imply additional branches. However, in the case of Lovelock theory of any order, there is only one branch, which is perturbative in $\a_m$, what can be easily seen by inverting the series (\ref{Lseries}). Now we are in position to consider static maximally symmetric solutions of the above theory allowing for an event horizon, electromagnetic field (expressed in the background electric charge of the object), and cosmological constant.

\section{Static black objects in the regularized Lovelock theory}\label{sec:regularized}
The generic $D$-dimensional static and maximally symmetric metric can be described by the following line element:
\begin{equation}\label{Lmetric}
  ds^2=-f(r)dt^2+\frac{1}{f(r)}dr^2 + r^2\gamma_{ij}dx^idx^j,
\end{equation}
where $d\Omega_n^2$ is a $(n=D-2)$-dimensional constant curvature space with a curvature $\kappa=-1,0,1$. The case $\kappa=1$ corresponds to the compact spherically symmetric solution, describing a black hole.

Starting from the Einstein-Lovelock-Maxwell theory and defining a new variable $\psi(r)$ via the following relation:
\begin{equation}\label{Lfdef}
f(r)=\kappa-r^2\,\psi(r),
\end{equation}
we obtain an algebraic equation \cite{Cai:2003kt}
\begin{equation}\label{MEq}
P[\psi(r)]=\frac{2\M}{r^{D-1}}-\frac{\Q^2}{r^{2(D-2)}}+\frac{2\Lambda}{(D-1)(D-2)}\,,
\end{equation}
where
\begin{equation}\label{Ppsi}
P[\psi]=\psi+\sum_{m=2}^{\m}\a_m\psi^m=\sum_{m=1}^{\m}\a_m\psi^m\,.
\end{equation}
When $\kappa=1$, the arbitrary constant $\M$ defines the asymptotic mass \cite{Myers:1988ze},
\begin{equation}\label{MADM}
  M=\frac{(D-2)\pi^{D/2-3/2}}{4\Gamma(D/2-1/2)}\M,
\end{equation}
and $\Q$ is the electric charge,so that the nonzero component of the electromagnetic strength tensor is
$$F^{tr}=\sqrt{\frac{(D-2)(D-3)}{2}}\frac{\Q}{r^{D-2}}.$$

For these particular solutions, as in the general theory above, for finite $\a_m$ all terms remain finite for $D\geq3$.

The $4D$ Eintein-Gauss-Bonnet theory ($\m=2$ in (\ref{Ppsi})) implies that
\begin{equation}
P[\psi]=\psi+\a_2\psi^2.
\end{equation}
Therefore, for $D=4$, Eq.~(\ref{MEq}) leads to two branches \cite{Fernandes:2020rpa},
$$
  f(r)=\kappa-\frac{r^2}{2\a_2}\left(-1\pm\sqrt{1+4\a_2\left(\frac{2\M}{r^3}-\frac{\Q^2}{r^4}+\frac{\Lambda}{3}\right)}\right).
$$
One of these branches, corresponding to the ``+'' sign, is perturbative in $\a_2$. For the ``-'' sign the metric function $f(r)$ goes to infinity when $\a_2 \to 0$. Notice that the above black-hole solution was also obtained in \cite{Cai:2009ua,Cognola} in a different context, when discussing quantum correction to entropy.

Higher-order Lovelock corrections result in more branches. However, there is always only one branch which is perturbative in $\a_m$. In particular, for $\m=3$ and $\a_3\geq\a_2^2/3$, we have
\begin{equation}
f(r)=1-\frac{\a_2r^2}{3\a_3}\left(A_+(r)-A_-(r)-1\right),
\end{equation}
where
\begin{eqnarray}\nonumber
A_{\pm}(r)&=&\sqrt[3]{\sqrt{F(r)^2+\left(\frac{3\a_3}{\a_2^2}-1\right)^3}\pm F(r)},\\\nonumber
F(r)&=&\frac{27\a_3^2}{2\a_2^3}\left(\frac{2\M}{r^3}-\frac{\Q^2}{r^4}+\frac{\Lambda}{3}\right)+\frac{9\a_3}{2\a_2^2}-1\,.
\end{eqnarray}
When $\a_3  <\a_2^2/3$, there are three real solutions to Eq.~(\ref{MEq}) resulting in three branches.

\section{Horizons and Hawking temperature}\label{sec:horizons}
Here we will consider the structure of the event horizons and the Hawking temperature for the above solutions.
For these purposes it is convenient to use the event horizon $r_H$ and the cosmological horizon $r_C>r_H$ instead of $\M$ and $\Lambda$. These units can be introduced as zeroes of the function $f(r)$. From (\ref{Lfdef}) we find that $\psi(r_H)=\kappa r_H^{-2}$, then Eqs.~(\ref{MEq}) and (\ref{Ppsi}) allow us to express the asymptotic mass as follows:
$$
  2\M=\frac{\Q^2}{r_H^{D-3}}-\frac{2\Lambda r_H^{D-1}}{(D-1)(D-2)}+\sum_{m=1}^{\m}\a_m \kappa^m r_H^{D-2m-1}.
$$
Using a similar equation which involves $r_C$,
$$
2\M=\frac{\Q^2}{r_C^{D-3}}-\frac{2\Lambda r_C^{D-1}}{(D-1)(D-2)}+\sum_{m=1}^{\m}\a_m \kappa^m r_C^{D-2m-1},
$$
one can obtain an expression for $\Lambda$ in an asymptotically de Sitter background,
\begin{eqnarray}\label{LdSdef}
  \frac{2\Lambda}{(D-1)(D-2)}&=&\Q^2\frac{r_C^{3-D}-r_H^{3-D}}{r_C^{D-1}-r_H^{D-1}}\\\nonumber&&+\sum_{m=1}^{\m}\a_m \kappa^m \frac{r_C^{D-2m-1}-r_H^{D-2m-1}}{r_C^{D-1}-r_H^{D-1}}.
\end{eqnarray}

The extremal value of $\Lambda$ can be obtained by taking the limit $r_C\to r_H$. Indeed, from the above equation we have
$$
  \frac{2\LE}{(D-2)}=-\frac{\Q^2(D-3)}{r_H^{2D-4}}+\sum_{m=1}^{\m}\a_m \frac{\kappa^m(D-2m-1)}{r_H^{2m}},
$$
which is finite again for any $D\geq3$.

When $D=4$ and $\kappa=1$, we obtain a particularly simple expression
\begin{equation}\label{LE4D}
  \LE=\frac{1}{r_H^2}-\frac{\Q^2+\a_2}{r_H^4}-\frac{3\a_3}{r_H^6}-\frac{5\a_4}{r_H^8}\ldots.
\end{equation}

In the asymptotically AdS background it is more convenient to measure the $\Lambda$-term in units of the AdS radius $R$, which is defined as \cite{Konoplya:2017lhs}
\begin{equation}
\psi(r\to\infty)=-\frac{1}{R^2}.
\end{equation}
Then, from Eq.~(\ref{MEq}) we obtain
\begin{equation}\label{LAdS}
  \frac{2\Lambda}{(D-1)(D-2)}=\sum_{m=1}^{\m}\frac{(-1)^m\a_m}{R^{2m}}.
\end{equation}

In addition,  one can check that surface gravity measured in these units obeys a simple formula \cite{Konoplya:2017lhs}
\begin{equation}\label{sg}
\frac{f'(r_H)}{2}=\frac{r_H(\LE-\Lambda)}{(D-2)P'[\psi(r_H)]},
\end{equation}
where
$$
P'[\psi(r_H)]=\frac{dP}{d\psi}[\psi(r_H)]=1+\sum_{m=2}^{\m}m\a_m\left(\frac{\kappa}{r_H^2}\right)^{m-1}.
$$
In particular, using (\ref{LE4D}) for $D=4$ and $\kappa=1$  we obtain an explicit expression for the Hawking temperature:
\begin{equation}\label{HawkingT}
T_H=\frac{f'(r_H)}{4\pi}=\frac{r_H}{4\pi}\frac{\dfrac{1}{r_H^2}-\dfrac{\Q^2+\a_2}{r_H^4}-\dfrac{3\a_3}{r_H^6}\ldots-\Lambda}{1+\dfrac{2\a_2}{r_H^2}+\dfrac{3\a_3}{r_H^4}\ldots},
\end{equation}
which coincides with (3.25) of \cite{Fernandes:2020rpa} for $\m=2$.

The value of the extreme charge $\QE$ as a function of the coupling constants $\a_i$, radius of the event horizon $r_H$ and the $\Lambda$-term can be obtained from the condition $T_H=0$,
$$\QE^2=r_H^2-\a_2-3\a_3 r_H^{-2}-5\a_4 r_H^{-4}\ldots-\Lambda r_H^4\,.$$

\section{Metric of the compact solution (black hole)}\label{sec:metric}
It was pointed out in \cite{Konoplya:2017lhs}, that for the compact solution ($\kappa=1$) of the perturbative branch $\psi(r)$ monotonically decreases within the following spans:
\begin{equation}\label{psispan}
\begin{array}{rclcc}
r_H^{-2}\geq&\psi(r)&\geq r_C^{-2} &\quad& \mbox{(de Sitter)},\\
r_H^{-2}\geq&\psi(r)&> 0 &\quad&  \mbox{(flat)},\\
r_H^{-2}\geq&\psi(r)&> -R^{-2} &\quad&  \mbox{(anti-de Sitter)}.
\end{array}
\end{equation}

This condition allows one to solve (\ref{MEq}) numerically for any given values of $D\geq3$, $\a_2$, $\a_3$, \ldots, $0<r_H<r_C\leq\infty$ (or $R>0$), and $\Q\leq\QE$. Numerical values of the derivatives of the function $\psi(r)$ can be further obtained from the differential equation,
$$P'[\psi(r)]\psi'(r)=-\frac{2(D-1)\M}{r^D}+\frac{2(D-2)\Q^2}{r^{2D-3}},$$
for any value of $r$.\footnote{The numerical method implemented in Wolfram~\emph{Mathematica}\textregistered{} can be downloaded from \url{https://arxiv.org/src/2003.07788/anc/LovelockBH.nb}.}

The perturbation equations of the charged Lovelock black hole were obtained in \cite{Takahashi:2012np}. Finite values for $\beta_m=m\a_m$ for $D\geq3$ correspond to regularization proposed in this paper. Although stability analysis is lacking in our case, it can be performed using the effective potentials obtained in \cite{Takahashi:2012np}, by taking the limit $D=4$.

\section{Conclusions}\label{sec:conclusions}
Here we suggested a natural generalization of the four-dimensional Einstein-Gauss-Bonnet theory of D.~Glavan and C.~Lin~\cite{Glavan:2019inb} to higher curvature terms in Lovelock form, that is, the four-dimensional Einstein-Lovelock theory. Further we considered static black objects in this theory which allow for a cosmological term and electric charge. The position of the event horizons and Hawking temperature is considered for these solutions. As the metric function cannot be represented in a closed form in the general case, the Mathematica\textregistered{} code which constructs the metric functions for every fixed set of parameters (such as the mass, charge, coupling constants, $\Lambda$-term, number of spacetime dimensions $D$) is developed. An important question which was beyond the scope of this paper is the stability of the above solutions against small perturbations of spacetime \cite{Konoplya:2011qq}. While the regime of small coupling constants should be stable, we anticipate eikonal instability in the regime of strong coupling, in a similar fashion with \cite{Konoplya:2020bxa}. Then, it would be interesting to study the quasinormal spectrum of the $4D$ Einstein-Lovelock theory in the regime of stability.

\bigskip

\emph{Note added: After the first version of this manuscript was uploaded into the arXiv a paper \cite{Casalino:2020kbt} (uploaded a day earlier) appeared, which has some (rather small) overlap with our research: In Sec.~IIIC of their work, a black hole for a particular case of the Lovelock theory $\a_3=\a_2^2/3$ was considered.}

\bigskip

\begin{acknowledgments}
The authors acknowledge the support of the grant 19-03950S of Czech Science Foundation (GAČR). This publication has been prepared with partial support of the ``RUDN University Program 5-100'' (R. K.).
\end{acknowledgments}

\end{document}